\documentclass[twoside,reqno]{bjp}
\usepackage{epsfig,cite}
\usepackage{graphics}
\usepackage{amssymb,amsmath}
\usepackage{times}
\begin{document}

\sloppy \raggedbottom

\setcounter{page}{1}

\title{\begin{center}
Age determination of possible binary open clusters NGC 2383/NGC 2384\\ and Pismis 6/Pismis 8
\end{center}}
\runningheads{Age determination of possible binary open clusters \ldots }{V.~Kopchev, G.~Petrov, P.~Nedialkov}

\begin{start}

\author{V.~Kopchev}{1}, \coauthor{G.~Petrov}{1},
\coauthor{P.~Nedialkov}{2}

\address{Institute of Astronomy, Bulgarian Academy of Sciences, 72 Tsarigradsko chaussee Blvd, 
1784 Sofia, Bulgaria}{1}

\address{Department of Astronomy, Sofia University, 5 James Bourcher Blvd, 1164 Sofia, Bulgaria}{2}

\received{\ldots}

\begin{Abstract}
Based on 2MASS J and Ks photometry for the open star clusters  NGC 2383, NGC 2384,
Pismis 6, Pismis 8 and using color magnitude diagrams with isochrones fit, we found an age of
$\log (\mbox{age})$ = 8.3 (200 $\pm$ 6 Myr) for NGC 2383 and $\log (\mbox{age})$ = 6.9 (8 $\pm$ 6 Myr) for NGC 2384.
For Pismis 6 and Pismis 8 we adopted a range of $\log (\mbox{age})$ = 6 - 7 (1 - 10 Myr).
Because they similar ages, Pismis 6 and Pismis 8 may have been formed in the same
Giant Molecular Cloud, and we concluded they are a good candidate for a binary system.
In the case of NGC 2383 and NGC 2384, because the big age difference found we conclude that  most probably 
they are born in different environments and as well are not physically connected.
\end{Abstract}


\end{start}

\section{Introduction}
Open clusters are very important objects in the study of stellar evolution
because their members are all of very similar ages and chemical composition. This way
the effects of other more subtle variables on the properties of stars are 
much more easily studied than they are for isolated stars.
The total number of open clusters known in our Galaxy is over 1600,
(see "New catalog of optically visible open clusters and candidates"  Dias et al. \cite{dias1}) 
of these the only well established double or binary cluster is NGC 869 and NGC 884
(known also as $h + \chi$ Persei), located at a distance of more than 2 kpc from the Sun.
The existence of other possible double clusters has been proposed earlier from Pavloskaya et al. \cite{pav2},
but not been seriously looked into. Subramaniam et al. \cite{sub3} examined existing catalogues of open clusters
and suggested 18 probable binary open star clusters.
The aim of this study is to determine the ages of two probable couples
NGC 2383 - NGC 2384 and Pismis 6 - Pismis 8. Our research is based on J and Ks photometry
from Two Micron All Sky Survey (2MASS project used two highly-automated 1.3-m telescopes, 
and provide all-sky photometry in J (1.25 microns), H (1.65 microns), and Ks (2.17 microns) bands).

\section{Earlier studies}
Table~{\ref{obs1} presents some data concerning previous investigations of objects under study.

\begin{table} [h]
\caption{\label{obs1} Earlier data for open clusters NGC 2383, NGC 2384, Pismis 6 and Pismis 8.
Denote: (V\&M) Vogt and Moffat; (S\&S) Subramaniam and Sagar;  (Fitz) Fitzegerald et al;
(B\&C) Battinelli and Capuzzo-Dolcetta; (F\&S) Forbes and Short.}
\begin{center}
\begin{tabular}{ccccc}
\hline\noalign {\smallskip}
Name & Citation & Distatnce     & Age & E(B-V)          \\
     &          &    pc         & Myr &  mag            \\
\hline\noalign {\smallskip}
NGC 2383 & V\&M & 1970          &     & 0.27            \\ 
.......	 & S\&S & 3340 $\pm$490 & 400 & 0.22 $\pm$0.05  \\
NGC 2384 & V\&M & 3280          &     & 0.29            \\
........ & S\&S & 2925 $\pm$430 & 20  & 0.28 $\pm$0.05  \\
Pismis 6 & V\&M & 1650          &     & 0.40            \\
........ & Fitz & 1700 $\pm$200 & 30  & 0.40            \\
........ & B\&C	&               & 32  & 0.41            \\
........ & F\&S	& 1850 $\pm$100 &  8  & 0.46 $\pm$0.04  \\
Pismis 8 & V\&M & 1420          &     & 0.74            \\
........ & Fitz & 1700 $\pm$200 & 30  & 0.74            \\
\hline
\end{tabular}
\end{center}
\end{table}

In the case of NGC2383 and NGC2384 Vogt and Moffat \cite{vogt4} from their photoelectric UBV measurements
concluded that despite both clusters appear close on the sky projection, 
because their big distances difference they are not physically connected. 
Subramaniam and Sagar \cite{sub5} presented first CCD photometry for NGC 2383 and NGC 2384, 
within the errors the distances are close enough to still match the selection criterion of Subramaniam et al. \cite{sub3},
but the large age difference indicates that the clusters did not form together from the same  Giant Molecular Cloud (GMC).

First studies of Pismis 6 and Pismis 8 are from Vogt and Moffat \cite{vogt6} using photoelectric photometry.
Besides distance measuremnts of both clusters Fitzegerald et al. \cite{fit7} found that they are
17 pc of each other in the plane of the sky, and that  their spectroscopic distance module are slightly larger than
photometric module (Fitzgerald et al. \cite{fit8}). The reddenings are compatible with the values derived by
Vogt and Moffat \cite{vogt6}. Fitzegerald suggested that Pismis 6 and Pismis 8 have an similar age of 30 Myr
and thus a common origin. For Pismis 6 Battinelli and Capuzzo-Dolcetta \cite{bat9}, Forbes and Short \cite{forb10} derived
32 Myr and 8 Myr respectively.

\section{Clusters data and age determinations}
Table~{\ref{obs2} present cluster parameters as given the catalog of Dias et al \cite{dias1}.

\begin{table} [h]
\caption{\label{obs2} Basic cluster parameters}
\begin{center}
\begin{tabular}{ccccc}
\hline\noalign {\smallskip}
Parameter            & NGC 2383   & NGC 2384     & Pismis 6    & Pismis 8    \\ 
\hline\noalign {\smallskip}
R.A.(2000).......... & 07:24:41.0 & 07:25:12.0   & 08:39:04.0  &  08:41:36.0 \\ 
Decl.(2000)......... & -20:56:42  & 3-21:01:24   & -46:13:36   &  -46:16:00  \\
Distance (pc)....... & 1655       & 2116         & 1668        &  1312       \\
Ang. diam (arcmin)   & 5          & 5            & 3           &  3          \\
E(B-V) (mag)...... ..& 0.213      & 0.255        & 0.380       &  0.706      \\ 
log(age)             & 7.167      & 6.904        & 7.283       &  7.427      \\
\hline
\end{tabular}
\end{center}
\end{table}

We used the photometry for the clusters from 2MASS using VizieR tool available at http://visier.u-strasbg.fr.
We made circular extractions centered on the coordinates for each clusters:
Apparent diameters on NGC 2383 and NGC 2384 is 5.0 arcmin and 3.0 arcmin for Pismis 6 and Pismis 8, 
for this we used an extraction radius of 5.0 arcmin and 3.0 arcmin respectively. 
Using the interstellar extinction law
$\frac{A_{J}}{A_{V}}$ = 0.282  and  $\frac{A_{K}}{A_{V}}$ = 0.112 from Rieke \& Lebofsky \cite{riek11}
and data for the distance and the reddening from \cite{sub5}:
\begin{flushleft}
$V-M_{V}$ = 13.3 $\pm$ 0.3 and $E_{B-V}$ = 0.22 $\pm$ 0.05 for NGC 2383\\ 
$V-M_{V}$ = 13.2 $\pm$ 0.3 and $E_{B-V}$ = 0.28 $\pm$ 0.05 for NGC 2384\\
\end{flushleft}
we derived reddening  $E_{J-K_{S}}$ = 0.12 $\pm$ 0.05  and $E_{J-K_{S}}$ = 0.15 $\pm$ 0.05 respectively. 
For other two clusters we using data from \cite{fit8}:
\begin{flushleft}
$V-M_{V}$ = 11.2 $\pm$ 0.2  and $E_{B-V}$ = 0.41 $\pm$ 0.06 for Pismis 6\\
$V-M_{V}$ = 11.0 $\pm$ 0.25 and $E_{B-V}$ = 0.76 $\pm$ 0.06 for Pismis 8\\
\end{flushleft}
and derived reddening $E_{J-K_{S}}$ = 0.25 $\pm$ 0.10 and  $E_{J-K_{S}}$ = 0.40 $\pm$ 0.10  respectively.\\
Colour-magnitude diagrams (CMDs) $M_{J}$ versus $(J-K_{S})_{0}$ for clusters are given on Figure \ref{fig1}.

\begin{figure}[h]
\centering{\epsfig{file=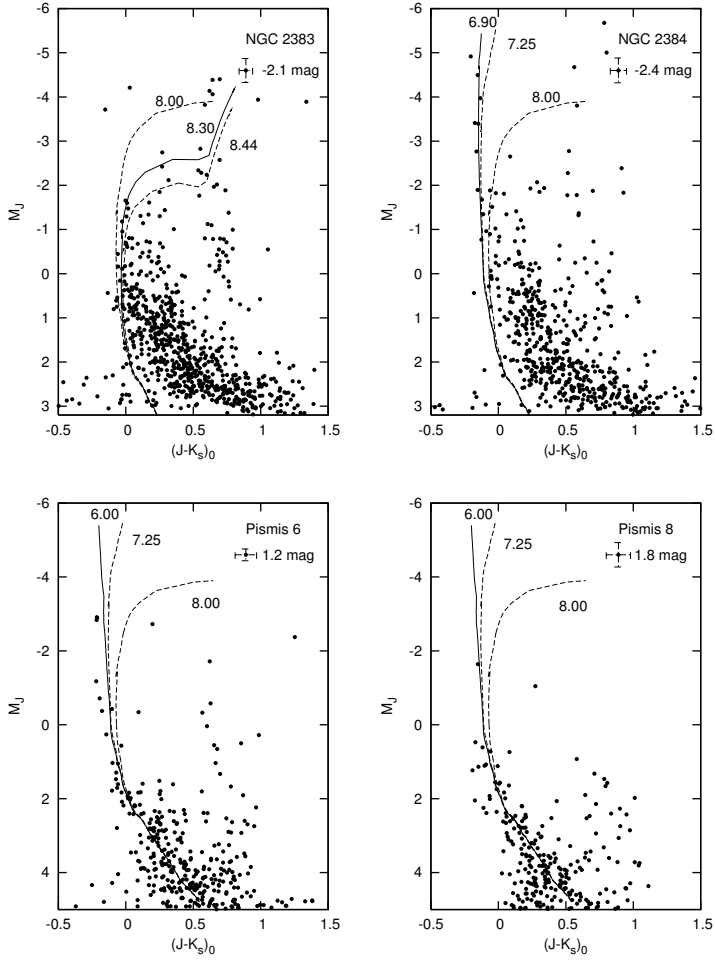}}
\caption{Colour-magnitude diagram for NGC 2383, NGC 2384, Pismis 6 and Pismis 8 with the best isochrone fit.
We have adopted an age of $\log (\mbox{age})$ = 8.3 (200 $\pm$ 6 Myr) for NGC 2383 and $\log (\mbox{age})$ = 6.9 (8 $\pm$ 6 Myr) for NGC 2384.
For Pismis 6 and Pismis 8 we adopted an range of $\log (\mbox{age})$ = 6 - 7 (1 - 10 Myr).}
\label{fig1}
\end{figure}

We determined the age of the clusters overplotting the best fitting isochrones on the CMDs. 
We have adopted an age of:
\begin{flushleft}
$\log (\mbox{age})$ = 8.3 (200 $\pm$ 6 Myr) for NGC 2383 and \\
$\log (\mbox{age})$ = 6.9 (8 $\pm$ 6 Myr) for NGC 2384
\end{flushleft}

For Pismis 6 and Pismis 8 we adopted an age range of $\log (\mbox{age})$ = 6 - 7 (1 - 10 Myr).
The isochrones are based on the stellar models of the Geneva group Schaerer D. et al.\cite{scha12},
with Z = 0.008 which corresponds to metallicity [Fe/H] = -0.3 dex. 
For NGC 2383 and NGC 2384 we determine less age from those given in \cite{sub5},
but for Pismis 6 and Pismis 8 our ages determination are comparable with \cite{fit8}, \cite{bat9}, and \cite{forb10}.

\section{Conclusions}
Using 2MASS  J and Ks photometry for the open star clusters  NGC 2383, NGC 2384, Pismis 6 and Pismis 8,
and fitting CMDs with isochrones based on the Geneva models, we found $\log (\mbox{age})$ = 8.3 (200 $\pm$ 6 Myr) for NGC 2383, $\log (\mbox{age})$ = 6.9 (8 $\pm$ 6 Myr) for NGC 2384, and range of $\log (\mbox{age})$ = 6 - 7 (1 - 10 Myr) for Pismis 6 and Pismis 8. Pismis 6 and Pismis 8 have similar age and may have been formed in the same GMC, and  we conclude that they are a good candidate for binary cluster. In contrast NGC 2383 and NGC 2384 have a big age range between, and may be not formed in the same GMC.

\section*{Acknowledgments}
Our work is partially supported by the grant F-1302/2003 of the Bulgarian NSF.
This publication makes use of data products from the Two Micron All Sky Survey,
which is a joint project of the University of Massachusetts and the Infrared Processing and Analysis Center/California Institute of Technology, funded by the National Aeronautics and Space Administration and the National Science Foundation.


\clearpage
\begin{thebibliography}{99}
\bibitem{dias1}
W. S. Dias, et al.(2002) {\em Astronomy and Astrophysics } {\bf 389} 871
\bibitem{pav2}
E. D. Pavlovskaya, A. A. Filippova (1989) {\em SvA } {\bf 33} 6
\bibitem{sub3}
A. Subramaniam, et al. (1995) {\em Astronomy and Astrophysics} {\bf 302} 86
\bibitem{vogt4}
N. Vogt, A. Moffat (1972) {\em Astronomy and Astrophysics Supplement Series} {\bf 7} 133
\bibitem{sub5}
A. Subramaniam, R. Sagar (1999) {\em Astronomical Journal} {\bf 117} 937
\bibitem{vogt6}
N. Vogt, A. Moffat (1973) {\em Astronomy and Astrophysics Supplement Series} {\bf 9} 97
\bibitem{fit7}
M. Fitzgerald, et al. (1979b) {\em Astronomy and Astrophysics Supplement Series} {\bf 37} 351
\bibitem{fit8}
M. Fitzgerald, et al. (1979a) {\em Astronomy and Astrophysics Supplement Series} {\bf 37} 345
\bibitem{bat9}
P. Battinelli, R. Capuzzo-Dolcetta (1991) {\em MNRAS} {\bf249} 76
\bibitem{forb10}
D. Forbes, S. Short (1994) {\em Astronomical Journal} {\bf 108} 594
\bibitem{riek11}
H.G. Rieke, M. J. Lebofsky (1985) {\em Astrophysical Journal} {\bf 288} 618
\bibitem{scha12}
D. Schaerer, et al. (1993) {\em Astronomy and Astrophysics Supplement Series} {\bf 98} 523
\end{thebibliography}
\end{document}